\documentclass[aps,reprint,longbibliography,showkeys,superscriptaddress]{revtex4-2}

\input{Misc/packages}
\begin{document}

\title{Shallow Electronic State Preparation for Quantum Chemistry with Quantum Monte Carlo Pre-Selection}

\author{Eline Welling}
\thanks{These authors contributed equally to this work.}
\affiliation{Yusuf Hamied Department of Chemistry, University of Cambridge, UK}
\affiliation{Fermioniq B.V., Amsterdam, NL}

\author{Lila Cadi Tazi}
\thanks{These authors contributed equally to this work.}
\affiliation{Yusuf Hamied Department of Chemistry, University of Cambridge, UK}

\author{Alex J. W. Thom}
\affiliation{Yusuf Hamied Department of Chemistry, University of Cambridge, UK}

\author{Maria-Andreea Filip}
\affiliation{Yusuf Hamied Department of Chemistry, University of Cambridge, UK}

\begin{abstract}
    Quantum computers hold great promise for molecular simulation, but noise remains a fundamental obstacle. We introduce a Quantum Monte Carlo (QMC) pre-screening procedure that constructs compact, physically motivated Givens rotation ans\"atze tailored to realistic quantum hardware. By identifying the most important wavefunction contributions early in a QMC simulation, we build circuits that are shallower than conventional alternatives while preserving number symmetry. 
    Benchmarked on Quantinuum System Model H1, QMC-prescreened Givens circuits outperform more complex ans\"atze under realistic noise conditions. The method offers a practical path toward chemical accuracy on quantum devices, by providing an adjustable trade‑off between expressivity and circuit depth to generate shallow circuits suited to current high‑noise devices, as well as deeper, more expressive circuits that can be deployed on future lower‑noise hardware.
\end{abstract}

\maketitle
\raggedbottom
\onecolumngrid

\section{Introduction}
In quantum chemistry, describing the electronic structure of molecules requires an exponentially large Hilbert space, which makes the classical simulation of many-body wavefunctions computationally demanding and unfeasible for large systems. Quantum computing provides a promising alternative, since a superposition of an exponential-scaling number of basis states can be encoded using only a linear-scaling number of qubits. As current quantum devices remain highly sensitive to noise, many practical approaches rely on hybrid quantum–classical algorithms. A prominent example is the variational quantum eigensolver (VQE) \cite{VQE_14}, which uses the variational principle to approximate the ground-state energy of a system. In this framework, a parameterised wavefunction ansatz is prepared on a quantum circuit; in principle, the more expressive this ansatz is, the better the resulting energy should be. However, in practice, more complex ans{\"a}tze lead to deeper circuits, which give rise to larger errors due to device noise, and prove challenging to optimise \cite{Marrero2021,Holmes2022,Ragone2023}.\\

One widely used VQE parameterisation is the unitary coupled cluster (UCC) \cite{Kutzelnigg1982,Kutzelnigg1983,Kutzelnigg1984,UCC89} ansatz. This wavefunction ansatz is relatively easy to implement, but leads to rather deep circuits, which are not without optimisation challenges \cite{Choy25}. There have been many attempts to reduce this depth \cite{Ryabinkin2018,Ryabinkin2020,Grimsley2019b,Tang2021,Burton2023}, including a recently proposed UCC Monte Carlo (UCCMC) \cite{Filip20, Filip22} approach, which classically screens amplitudes and only encodes the most significant contributions of the wavefunction into the ansatz. \\

Alternative ans{\"a}tze have been developed that focus on compatibility with hardware, such as the hardware-efficient ansatz \cite{HEA17}. However, these can suffer from the barren plateau problem \cite{Barren_plateau2018}, partly due to the many unphysical states they produce, which hinders VQE convergence. Effort has gone into ways to only create states which obey the symmetries of the system \cite{Burton2024,Motta2023}, of which we note in particular the use of Givens rotations \cite{Arrazola22}.\\

In this work, we present a Monte Carlo pre-screening method for a Givens rotation based wavefunction parameterisation, which is inherently more hardware-efficient than UCC. The Givens ansatz has been previously proposed in the so-called qubit coupled cluster (QCC) methods \cite{Yordanov2020,Xia2020,Yordanov2021,Xie2022}, and has recently been shown to lead to significant circuit depth reduction for molecular examples \cite{Rajan2025}. \\

The resulting screened ans{\"a}tze for a range of molecules were implemented on Quantinuum System Model H1, with results showing that, in the presence of noise, screened Givens ans{\"a}tze significantly outperform UCC circuits for the same number of parameters. The screening approach allows for a simple mechanism to trade off potential accuracy for circuit depth, in order to generate optimal results given particular noise behaviour. \\

\section{Results}

\subsection{Prescreened QMC-Givens Ansatz}

In recent years, Hilbert-space Quantum Monte Carlo (QMC) algorithms \cite{Booth2009,Thom2010,Filip20} have been developed as highly accurate alternatives to conventional electronic structure methods, taking advantage of the sparsity of the wavefunction to decrease memory (and often computational) requirements relative to comparable deterministic techniques. They have been used to obtain benchmark results for challenging chemical systems \cite{Booth2011,Ghanem2020,Eriksen2020}. They have also been employed to select important contributions to the wavefunction, to use in methods such as the CC(P,Q) family of approaches\cite{Magoulas2023,Magoulas2023b}, which approximates high-level coupled cluster energies by including carefully selected highly-excited amplitudes in a low-level coupled cluster calculation. \\

QMC approaches are usually based on an iterative implementation of a linearisation of the imaginary-time-evolution operator to project onto the ground state. The wavefunction $\Psi$ is given by some ansatz determined by a vector of parameters, $\boldsymbol{\theta}$, and written $\Psi(\boldsymbol{\theta})$.  This evolves in imaginary time $\beta$ to the ground-state wavefunction by evolving its parameters $\boldsymbol{\theta}(\beta)$.  We may thus write $\Psi(\beta)=\Psi(\boldsymbol{\theta}(\beta))$, and evolve
\begin{equation}
    \ket{\Psi(\beta + \Delta\beta)} = (1 - \Delta\beta (\hat H - S)) \ket{\Psi(\beta)},
\end{equation}
where $\hat H$ is the Hamiltonian of the system of interest, $S$ is a population-control parameter which will normally converge to the ground-state energy and $\Delta \beta$ is the (real) step in imaginary time. This can be translated directly into an update equation for the ansatz parameters, provided that for each parameter $\theta_i$ there exists an independent configuration (typically a Slater determinant) $\ket{\Phi_i}$, such that
\begin{equation}
    \braket{\Phi_i|\Psi(\beta)} = \theta_i(\beta) + \mathcal{O}(\theta^2),
    \label{eq:linear}
\end{equation}
in which case
\begin{equation}
    \theta_i(\beta + \Delta\beta) = \theta_i(\beta) - \Delta\beta \braket{\Phi_i|\hat H - S|\Psi(\beta)}.
    \label{eq:qmc_update}
\end{equation}
\textcite{Filip20} have shown that UCCMC can be effectively used to select UCC parameters for simplified VQE calculations \cite{Filip22}, in a methodology we will refer to here as QMC-UCC. This approach recovers most of the correlation energy in molecular systems for a fraction of the cost. However, due to the structure of the excitation operators, even screened UCC leads to deep quantum circuits. We therefore propose extending this approach to a shallower parametrisation based on Givens rotations.\\

The controlled Givens ansatz \cite{Arrazola22} (see \cref{sec:controlled_givens}) gives rise to a wavefunction of the form
\begin{equation}
   \ket{\Psi(\boldsymbol{\theta})} = \prod_{k=1}^{N} \cos(\frac{\theta_k}{2}) \ket{\Phi_0} + \sum_{i=1}^{N} \sin(\frac{\theta_{i}}{2}) \prod_{j=1}^{i-1} \cos(\frac{\theta_j}{2}) \ket{\Phi_{i}}, 
\end{equation}
which has the form of \Cref{eq:linear}. Therefore, the QMC algorithm can be straightforwardly adapted to this ansatz. We will refer to both the Monte Carlo evolution of the Givens ansatz and the resulting screened parametrisation as QMC-Givens. \\

In the QMC algorithm, the parameters $\theta_i$ are determined by a (typically integer) population of walkers which may reside on a site for each parameter.  The number of walkers on the site corresponding to $\theta_i$ is denoted $N_i$ and we can recover $\theta_i = N_i/N_0$, where $N_0$ is a reference population for normalization, corresponding to $\braket{\Phi_0|\Psi(\beta)}$. In the following, we will refer to the stochastic rounding of a number $r$, which is given by $\mathrm{SR}(r) = \lfloor r \rfloor + \text{Bernoulli}(r-\lfloor r \rfloor)$, where $\text{Bernoulli}(x)$ is a Bernoulli trial with probability $x$. \Cref{eq:qmc_update} can be implemented stochastically by three processes:
\begin{enumerate}
    \item \textbf{Spawning} corresponding to the off-diagonal terms in \Cref{eq:qmc_update}, $\Delta \beta C_j N_0 H_{ij}$, where $j$ refers to one of the determinants making up $\Psi(\boldsymbol{\theta})$. For each \textit{determinant}, $\ket{\Phi_j}$, in the wavefunction, $C_j=\braket{\Phi_j|\Phi(\beta)}$ is evaluated, and a spawn is attempted $\mathrm{SR}(C_j N_0)$ times.  Each spawn consists of randomly selecting a particular $\theta_i$. A number of particles 
    \begin{equation}
        N_s = -\mathrm{sgn}(C_j)\mathrm{SR}(\Delta\beta H_{ij}/p_\mathrm{gen}(i|j))
    \end{equation}
    is spawned on $\theta_i$, where $p_\mathrm{gen}(i|j)$ is the probability of selecting $\theta_i$ as the spawning target. In our current implementation, this is uniform.

    \item \textbf{Death} corresponding to the diagonal terms in \Cref{eq:qmc_update}, $\Delta \beta C_i N_0 (H_{ii}-S)$. For each \textit{determinant}, $\ket{\Phi_i}$, in the wavefunction, death is attempted $\mathrm{SR}(C_i N_0)$ times. A number of particles 
    \begin{equation}
        N_d = -\mathrm{sgn}(C_i)\mathrm{SR}(\Delta\beta (H_{ii}-S))
    \end{equation} is generated on $\theta_i$.

    \item \textbf{Annihilation} Positive and negative contributions to $N_i$ cancel out, preventing the development of oppositely signed copies of the wavefunction simultaneously.
\end{enumerate}

A simple implementation of QMC-Givens is available in Ref. \cite{GivensQMC}. \\

We note that the uncontrolled Givens ansatz (see \cref{sec:unc_givens}) is also of a form amenable to a QMC treatment, and the necessary approach is similar to the trotterised or disentangled UCCMC method \cite{Filip20}. However, we expect that, similarly to UCC, the Givens ansatz will perform better in regimes where the optimal values for the parameters $\theta_i$ are small. In this case, the controlled Givens ansatz corresponds to a linear approximation of the uncontrolled Givens ansatz with the same gates. We therefore expect that parameters from a controlled QMC-Givens run should be good indicators for the values of parameters in an equivalent uncontrolled QMC-Givens calculation and use the former as our screening function, due to its relative ease of classical implementation.\\

The QMC representation will be naturally sparse, with parameters of small magnitude likely to be populated later in a calculation or possibly have zero population on many individual iterations. In UCCMC pre-screening, the population on particular parameters at the end of a short, likely unconverged UCCMC calculation was used to decide whether to include them in the VQE ansatz. Here, we implement a simpler heuristic, where we look at the first time a determinant becomes populated in a calculation as an indicator of its final magnitude in the wavefunction. For an $n$-parameter VQE optimisation, we then select the first $n$ parameters to be spawned to in the QMC-Givens run. In the following section, we consider the validity of this approach.\\

We studied the performance of the QMC-Givens method for three molecular systems at equilibrium geometries: lithium hydride (\ce{LiH} with a bond length of $1.40$\AA), beryllium hydride (\ce{BeH2} in a linear configuration with H–Be bond distances of $1.20$\AA), and nitrogen (\ce{N2} with a bond length of $1.098$\AA). The STO-3G basis set was used for all three molecules, with a restriction to $s$ orbitals only for lithium and beryllium. For the N$_2$ molecule, the core $1s$ orbitals are frozen. Aside from \Cref{fig:correlationplot_N2}, which shows the full excitation space, all results use a Givens ansatz constrained to single and double excitations.

\subsection{QMC prescreening performance}\label{sec:results_qmc}

To assess the QMC prescreening method, we examined the relationship between each parameter's first spawning iteration and its final contribution to the wavefunction, to determine whether early-simulation data can predict which parameters will carry the largest final amplitudes. For each parameter, we recorded the median spawning iteration across ten independent runs and compared these against the converged final coefficients.

\begin{table}[H]
\centering
\caption{\centering Summary of correlation statistics for 10 independent QMC-Givens 
runs per system. Spearman correlations are computed between spawning 
iteration and final determinant coefficients.}
\label{tab:correlations}
\begin{tabular}{P{18mm} P{25mm} P{25mm} P{25mm}}
\toprule
\textbf{System} & \textbf{$\rho_{\text{10runs}}$} & \textbf{p-value} & \textbf{Mean $\rho$ per run} \\
\midrule
LiH      & $-0.783$ & ${\phantom{<}}1\times10^{-2}$    & $-0.684 \pm 0.133$ \\
BeH$_2$  & $-0.939$ & ${\phantom{<}}5\times10^{-5}$    & $-0.899 \pm 0.061$ \\
N$_2$    & $-0.855$ & ${<}1\times10^{-9}$ & $-0.717 \pm 0.021$ \\
\bottomrule
\end{tabular}
\end{table}

We quantified this relationship using the Spearman rank correlation coefficient \cite{Spearman1904} between the median spawning iteration and the final determinant coefficients, defined as
\begin{equation}
    \rho = \frac{\text{cov}[R[t]R[\theta]]}{\sigma[R[t]]\sigma[R[\theta]]},  
\end{equation}
where $R[t]$ and $R[\theta]$ are the rank variables corresponding to the average spawning iteration and the final parameter value, $\text{cov}[X,Y]$ denotes the covariance between two random variables and $\sigma[X]$ is the standard deviation of random variable $X$. \\

Table \ref{tab:correlations} shows a strong negative correlation across a range of molecules, confirming that parameters spawned earlier tend to have larger final amplitudes. The p-value quantifies the probability of observing such a correlation in the absence of a true relationship; the elevated value for the smallest system reflects its fewer determinants and correspondingly lower statistical significance. For the larger N$_2$ system, both single-run and combined analyses yield strong correlations, with multi-run aggregation producing a more reliable result. Since independent runs can be performed in parallel, this offers an efficient route to improved prescreening performance.

\begin{figure}[ht]
    \centering
    \includegraphics[width=0.45\linewidth]{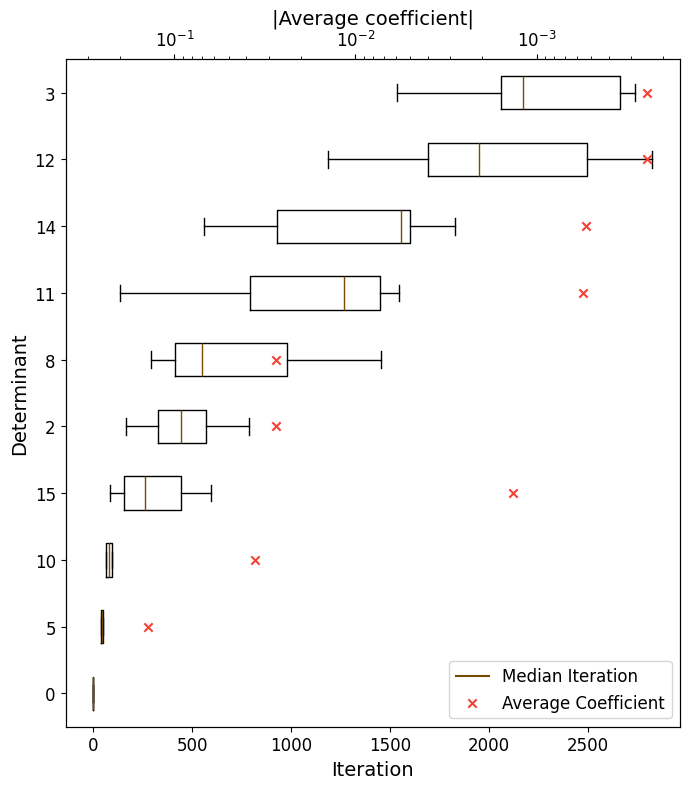}
    \caption{BeH$_2$: Boxplots show the distribution of iterations at which each determinant is first spawned across 10 independent QMC-Givens runs of BeH$_2$. Medians are indicated by vertical lines, with boxes spanning the interquartile range. Crosses (×) mark the absolute value of the average coefficient of each determinant at convergence, averaged over the same 10 runs. Determinants are given an index and then ordered by spawning iteration in the plot; those spawned earlier tend to carry larger weights in the converged wavefunction.}
    \label{fig:boxplot_BeH2}
\end{figure}

\begin{figure}[ht]
    \centering
    \includegraphics[width=0.45\linewidth]{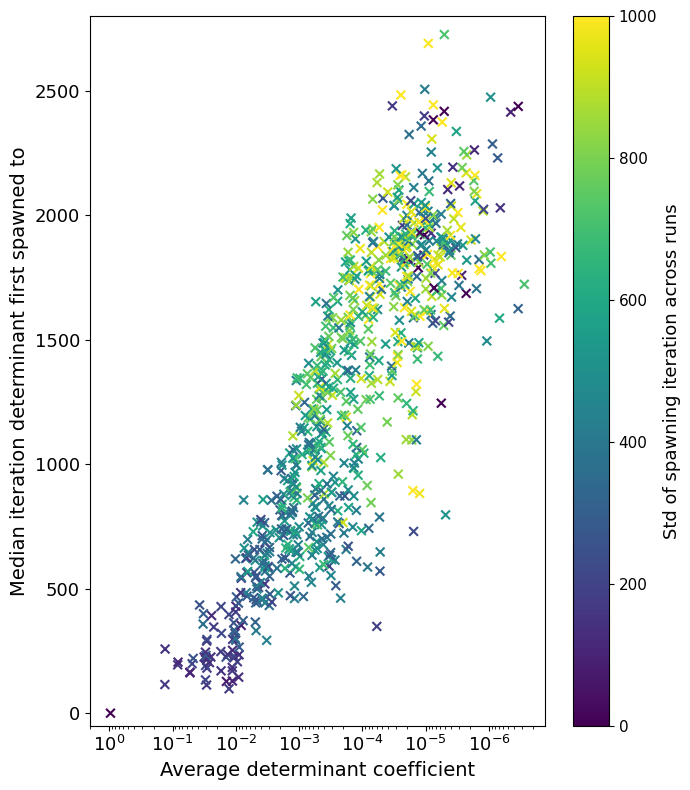}
    \caption{N$_2$: Average determinant coefficient magnitude versus median spawning iteration across 10 independent QMC-Givens runs. Each cross marks one determinant, with color indicating the standard deviation of its spawning iteration across the 10 runs. Although individual spawning events show substantial run-to-run variation, the overall ordering is reproducible across runs: determinants with larger converged coefficients are consistently spawned earlier in the sampling (see Table~\ref{tab:correlations}).}
    \label{fig:correlationplot_N2}
\end{figure}

This behaviour is further visualized in Fig. \ref{fig:boxplot_BeH2} for BeH$_2$, which shows a clear trend in first spawning iterations across runs, but also a noticeable spread, reinforcing that combining multiple runs improves the consistency of early determinant identification. Fig.~\ref{fig:correlationplot_N2} shows the relationship between first spawning iteration and average final parameter amplitude across 10 runs for N$_2$. There is a clear monotonic trend, demonstrating that short QMC runs can identify the most important determinants even before reaching the population plateau necessary for convergence.\\

\begin{figure}[ht]
    \centering
    \includegraphics[width=0.6\linewidth]{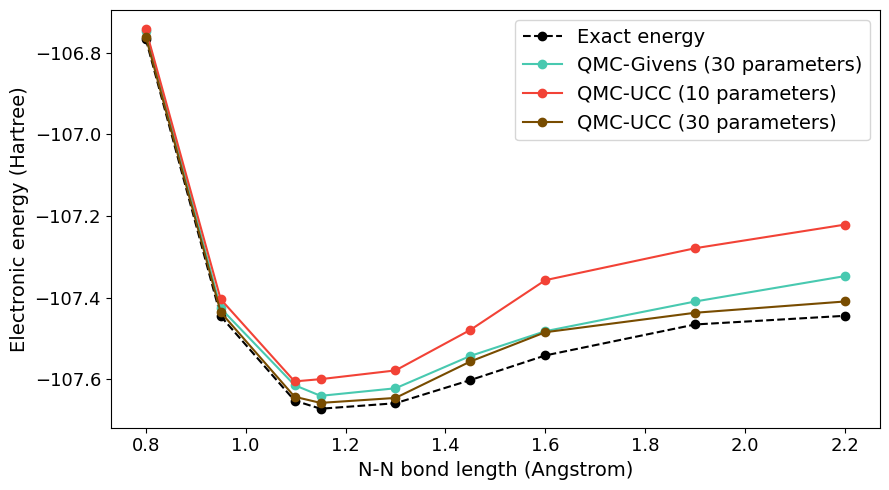}
    \caption{N$_2$ binding curve. Solid curves show statevector VQE energies for the QMC-Givens ansatz (30 excitation parameters) and QMC-UCCSD ansätze with 10 and 30 parameters. The black dashed curve is the exact ground-state energy obtained by full diagonalization of the electronic Hamiltonian in this basis. We note that as this is a stochastic procedure, some variation in the results is possible.}
    \label{fig:bindingcurve_N2}
\end{figure}

While short QMC runs can predict the important contributions to the final wavefunction, we must also assess the error incurred by including only a subset of determinants, particularly as molecular correlation increases. To test performance across different correlation strengths, we constructed a statevector VQE binding curve for N$_2$ using 30-parameter Givens circuits built from single and double excitation gates selected via QMC prescreening. At each bond length, the 30 determinants with the earliest median spawning iterations from short QMC-Givens runs were included. We then compared these to UCCSD circuits with the same number of parameters screened using UCCMC as described in Ref. \cite{Filip22}, with results shown in \Cref{fig:bindingcurve_N2}. The QMC prescreening performs reasonably well even at longer bond lengths where the wavefunction becomes increasingly multi-reference, though the discrepancy between the Givens ansatz and UCCSD grows with correlation strength. We explore the reasons for this further in \cref{sec:results_vqe}. 

\begin{figure*}[t]
    \centering

    \begin{minipage}{0.32\textwidth}\centering $\textbf{\ce{LiH}}$ \textbf{- 6 qubits}\end{minipage}\hfill
    \begin{minipage}{0.32\textwidth}\centering $\textbf{\ce{BeH2}}$ \textbf{- 8 qubits}\end{minipage}\hfill
    \begin{minipage}{0.32\textwidth}\centering $\textbf{\ce{N2}}$ \textbf{- 16 qubits}\end{minipage}

    \vspace{2mm}

    \includegraphics[width=0.32\textwidth]{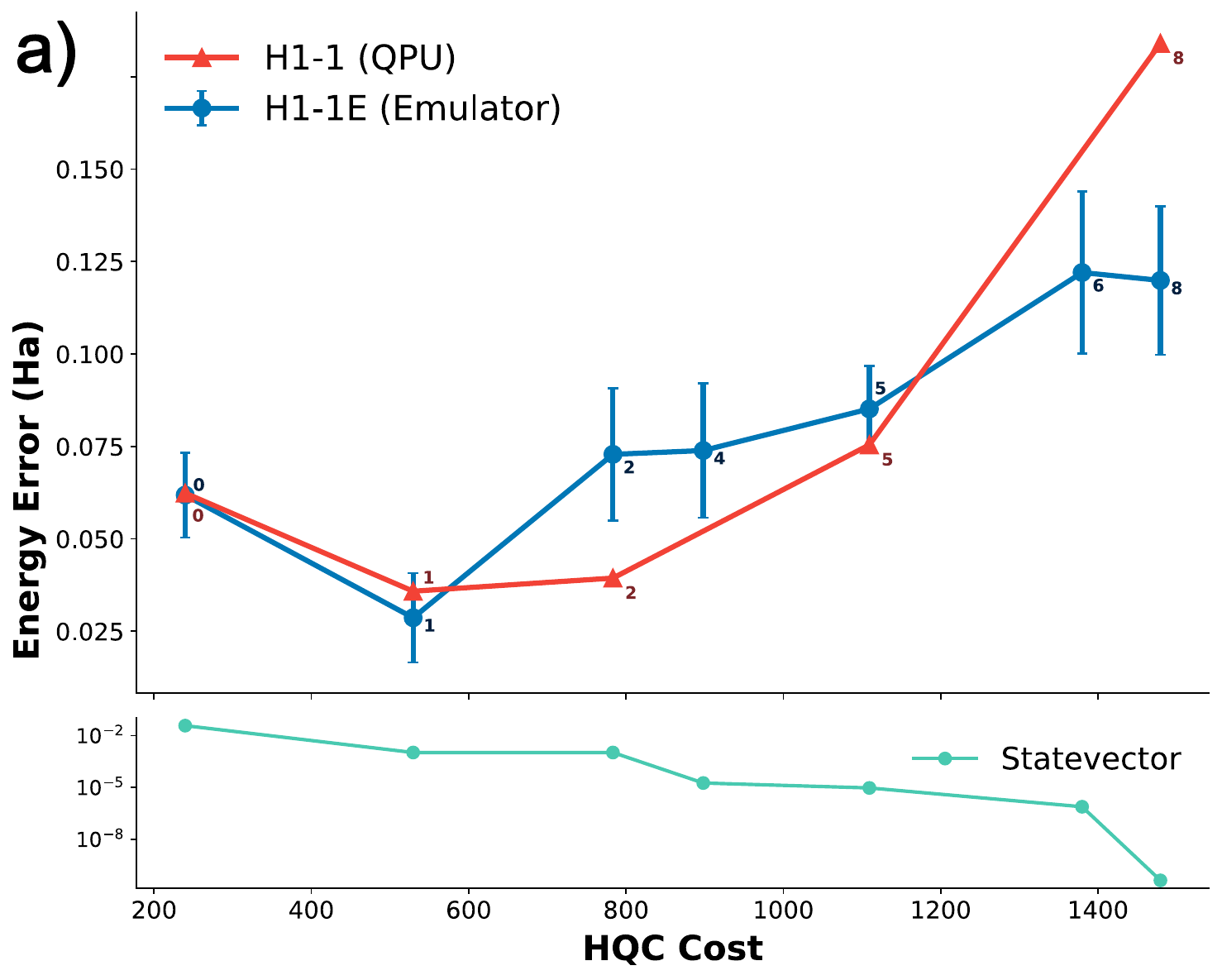}\hfill
    \includegraphics[width=0.32\textwidth]{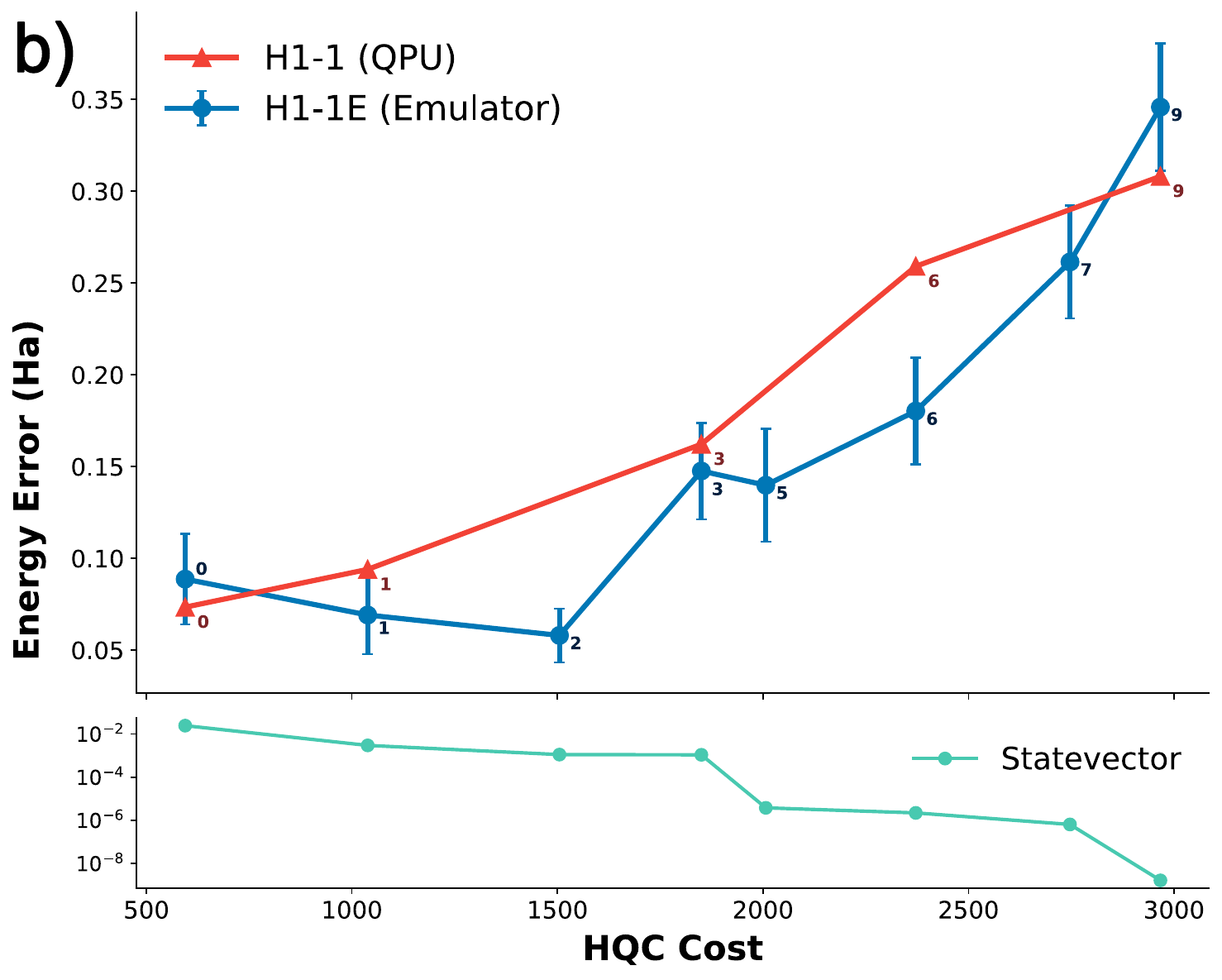}\hfill
    \includegraphics[width=0.32\textwidth]{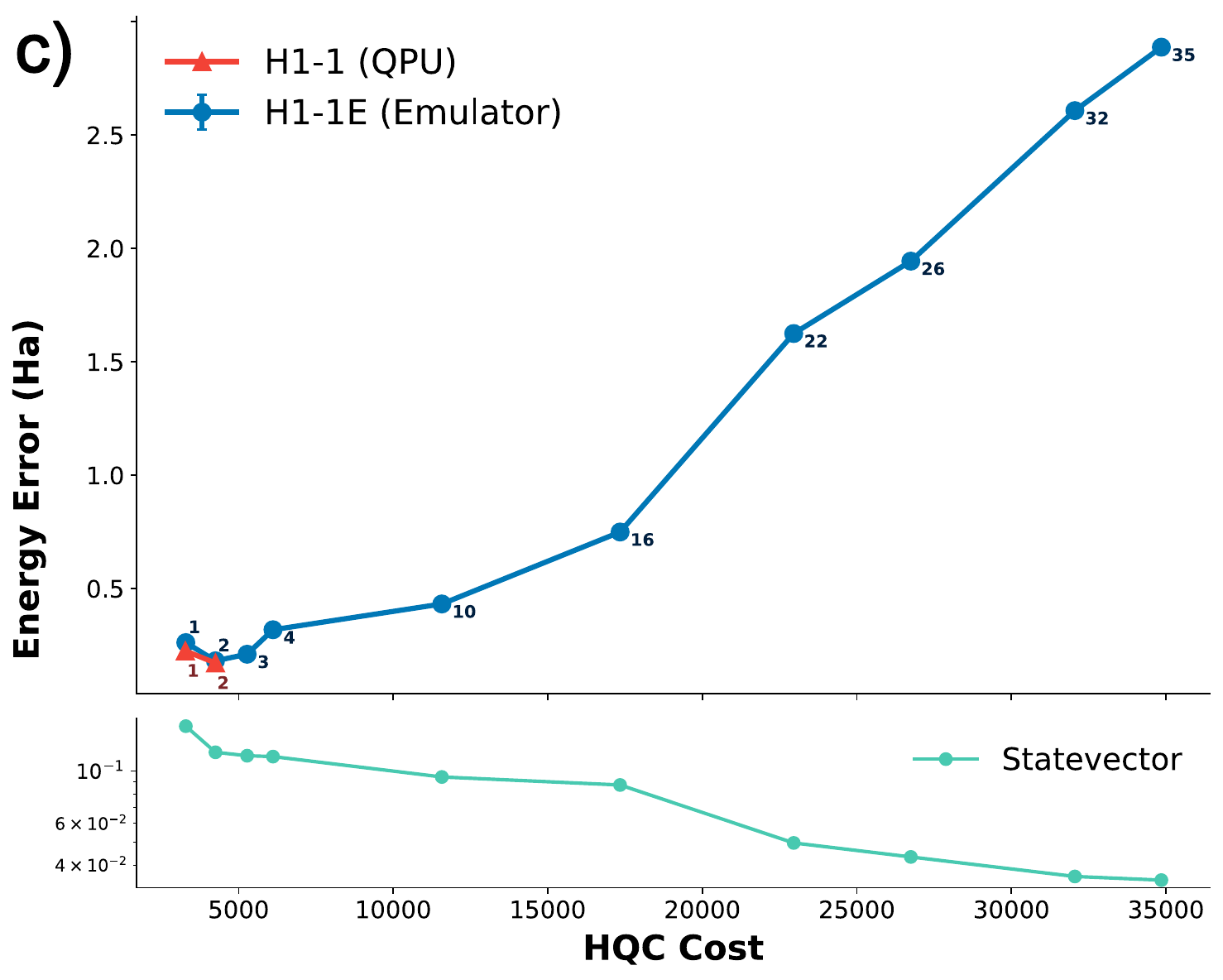}

    \vspace{2mm}

    \includegraphics[width=0.32\textwidth]{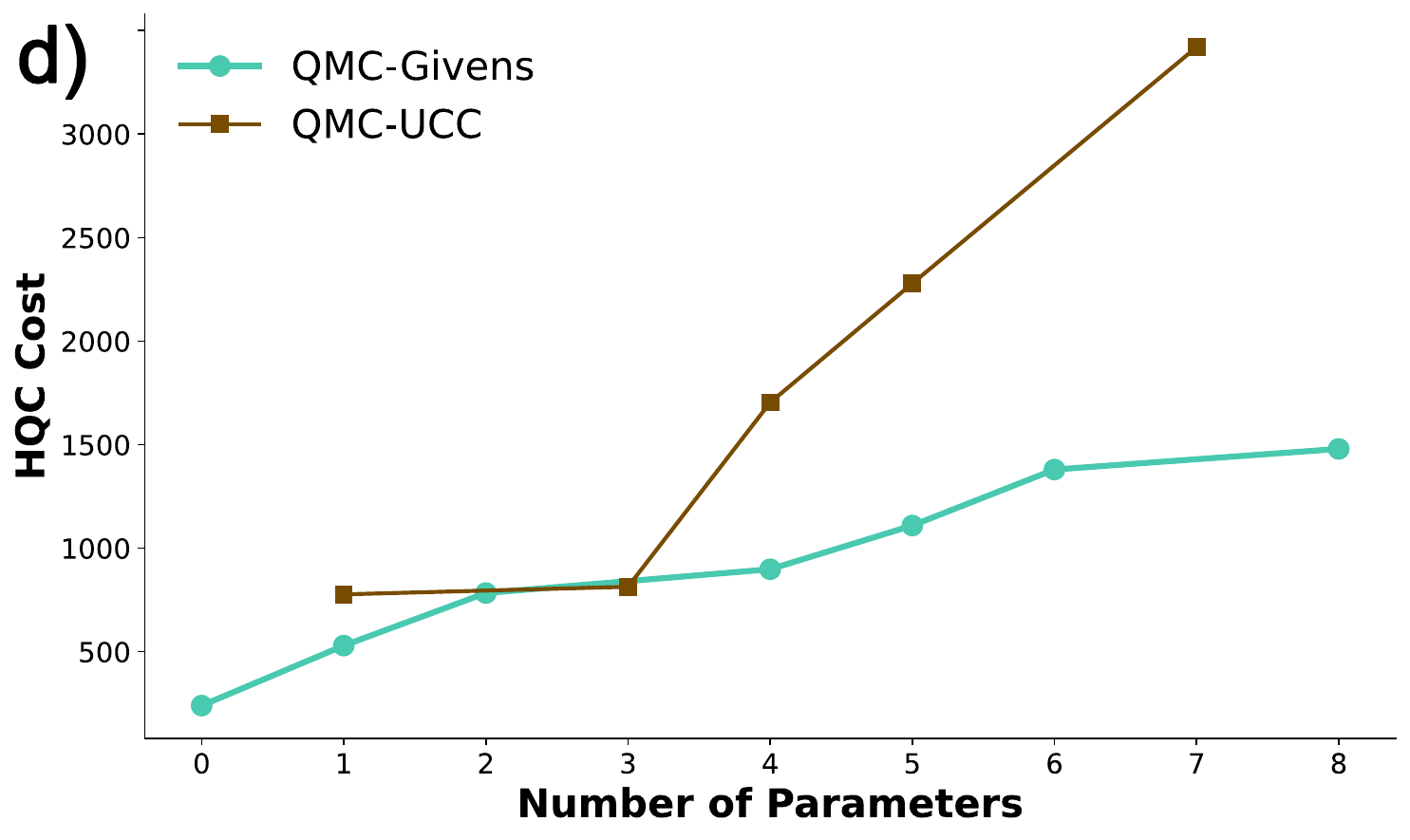}\hfill
    \includegraphics[width=0.32\textwidth]{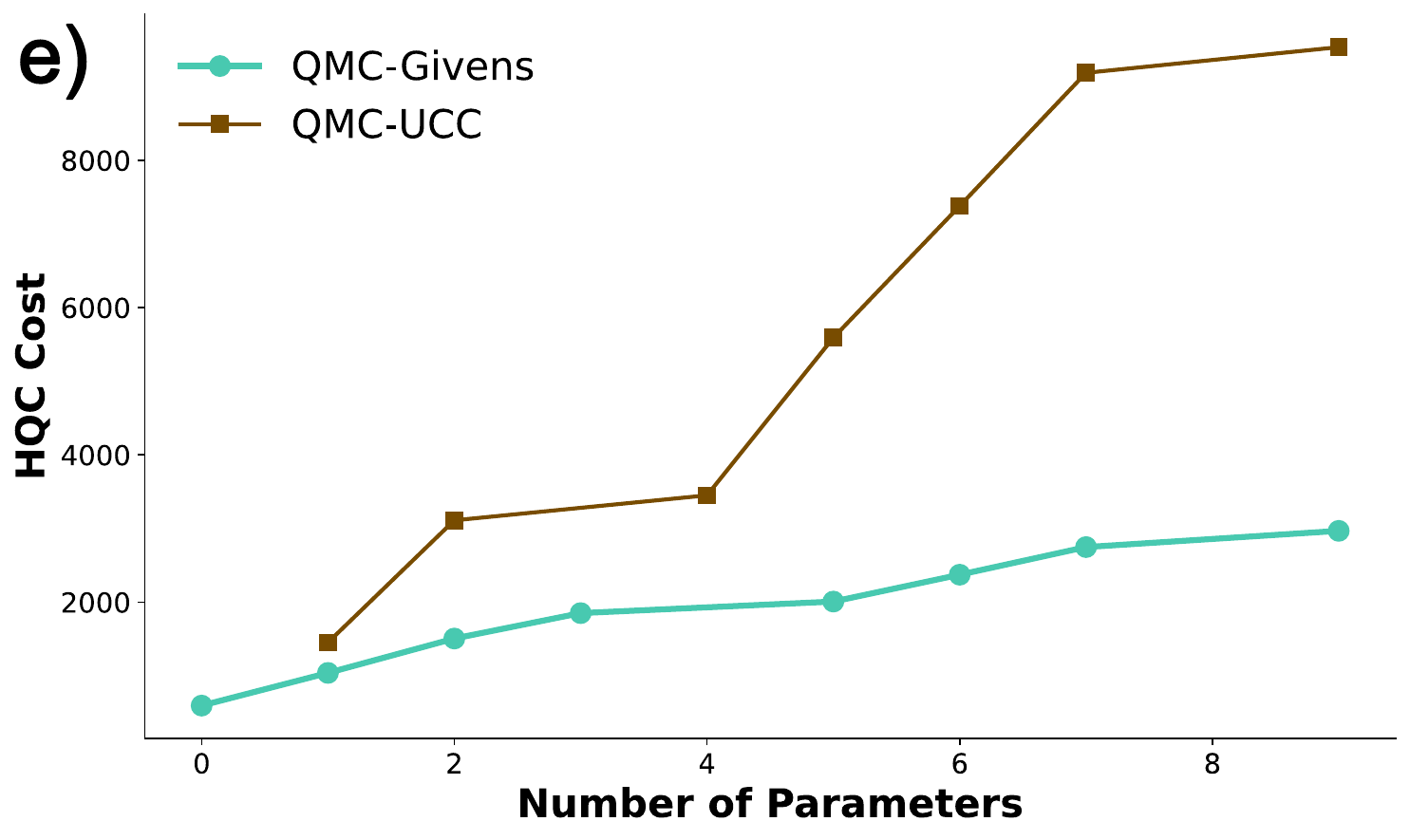}\hfill
    \includegraphics[width=0.32\textwidth]{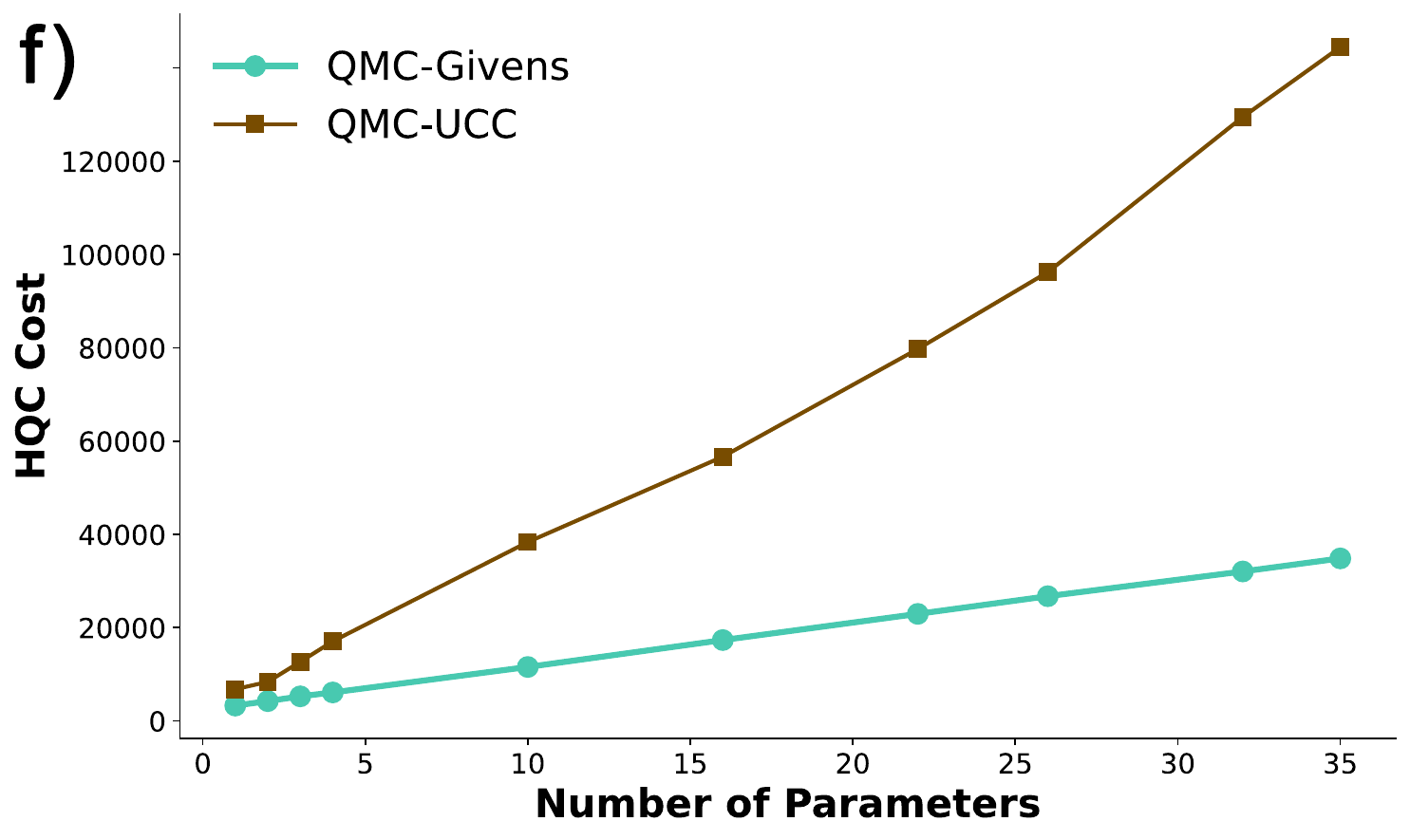}

    \vspace{2mm}

    \includegraphics[width=0.32\textwidth]{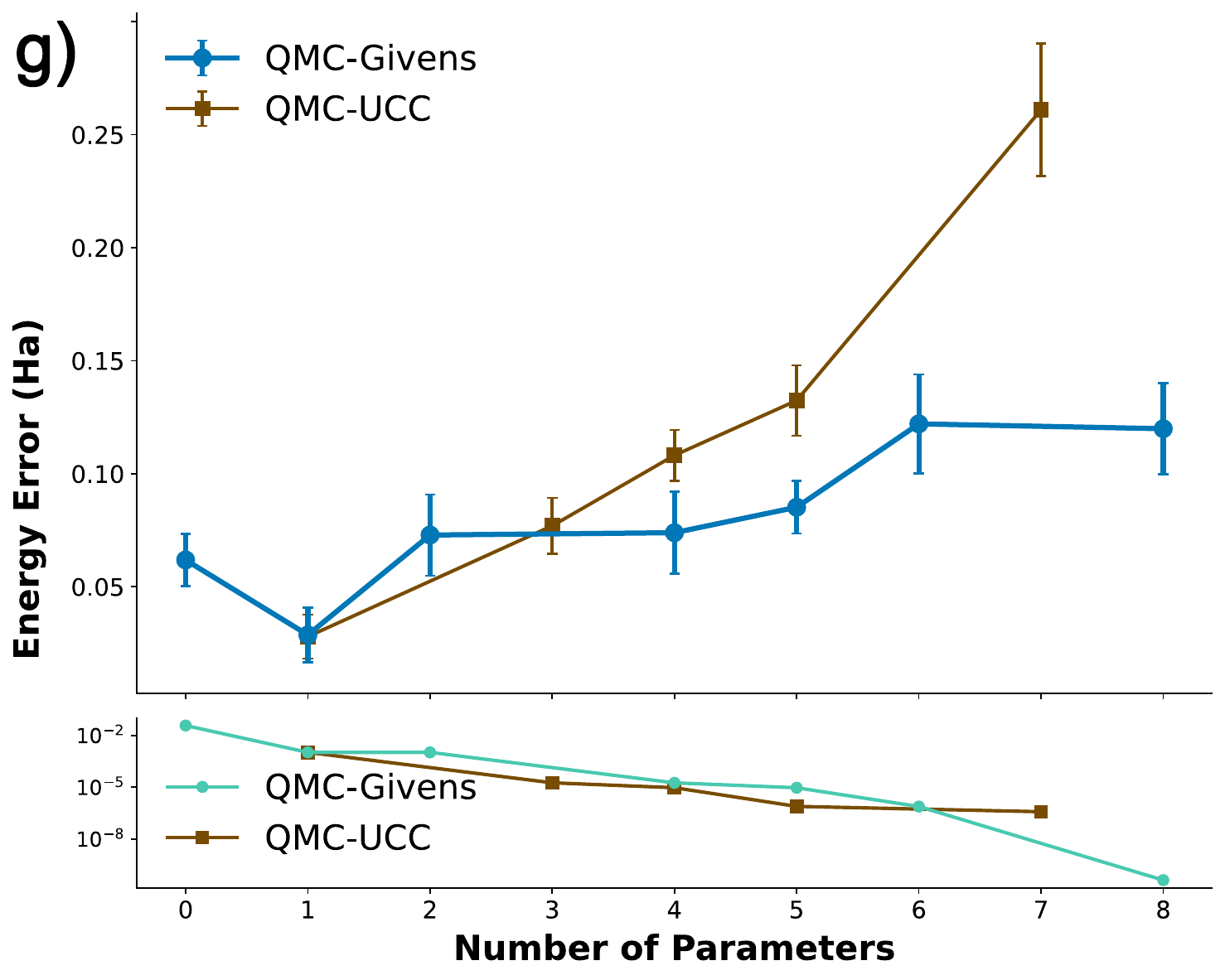}\hfill
    \includegraphics[width=0.32\textwidth]{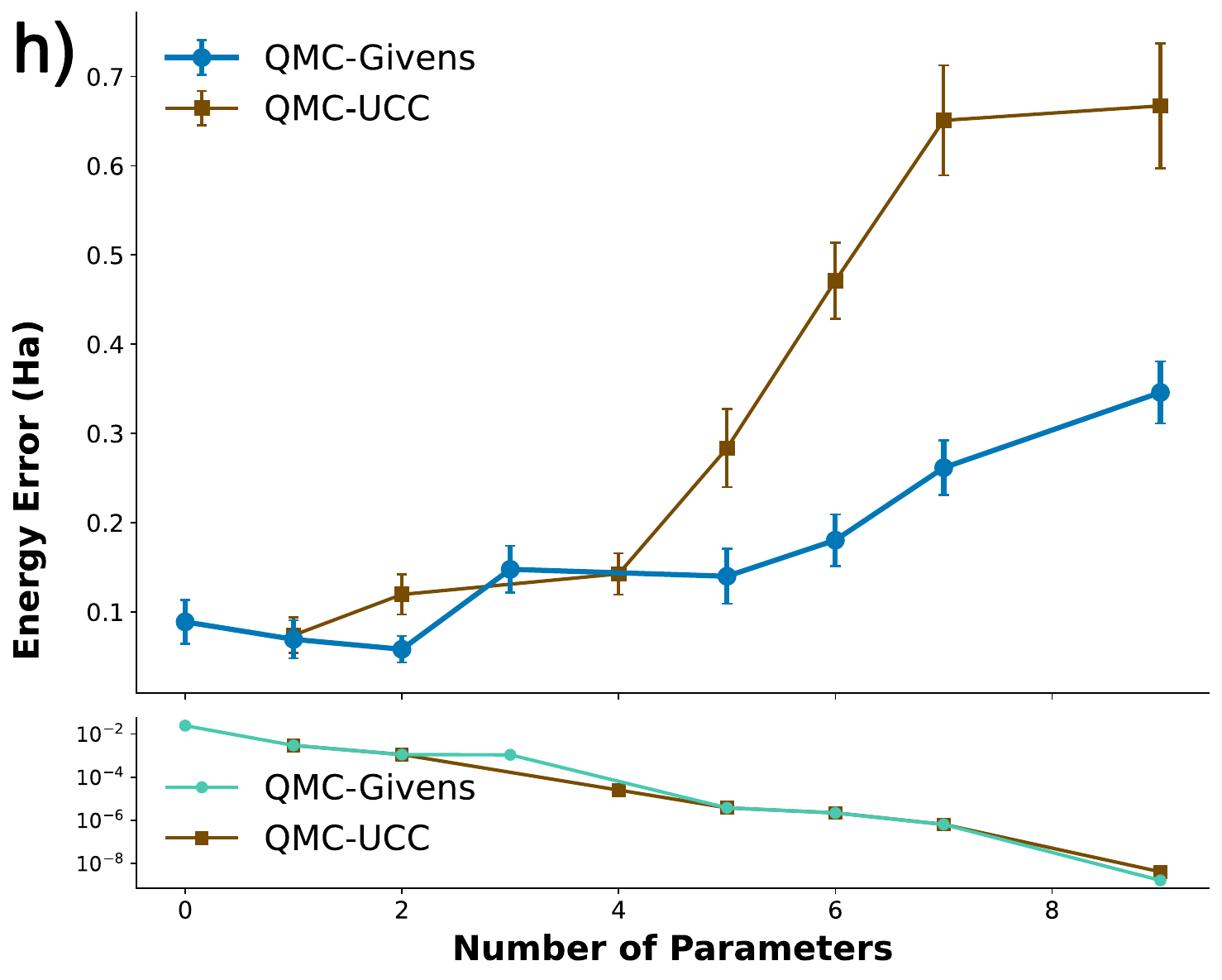}\hfill
    \includegraphics[width=0.32\textwidth]{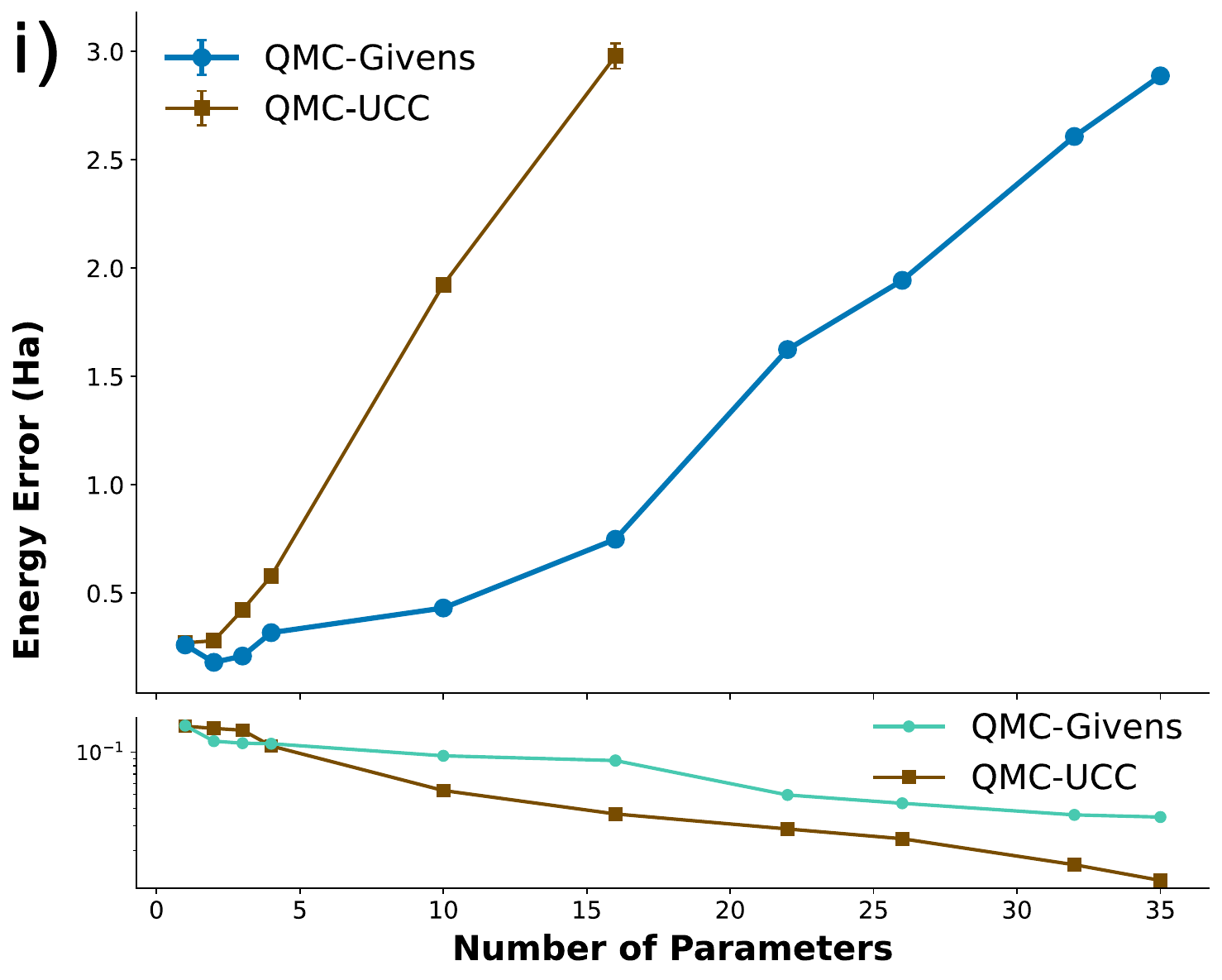}

\caption{
Performance of the QMC-Givens and QMC-UCC ansätze for $\ce{LiH}$, $\ce{BeH2}$, and $\ce{N2}$.
Columns correspond to molecular systems, as indicated at the top.
All energies were computed using 1000 shots; where applicable, error bars are estimated from 10 independent samples.
\textbf{a) - c):} Energy accuracy as a function of HQC cost for increasing circuit sizes, shown for H1 hardware, the H1-1E noisy emulator, and ideal statevector simulations.
\textbf{d) - e):} HQC cost versus number of variational parameters for QMC-Givens and QMC-UCC circuits.
\textbf{f) - i):} Energy accuracy versus number of parameters, comparing QMC-Givens and QMC-UCC on the H1-1E emulator (upper curves) and the statevector simulator (lower curves).}
    \label{fig:vqe_figures}
\end{figure*}

\subsection{Quantum Hardware and Simulation Results}\label{sec:results_vqe}

\subsubsection{QMC-Givens ansatz on quantum hardware and simulators}

For each of the three molecules considered, pre-screened circuits with varying numbers of parameters were prepared. The variational parameters of each ansatz circuit were then optimised 
using a VQE algorithm on an ideal classical simulator. Pre-optimised circuits were  executed on quantum hardware (Quantinuum H1) or a emulator (H1-1E) to compute energy expectation values. This methodology was chosen because of the limited quantum resources available and to serve as a reference for the best achievable performance of each ansatz at optimal angles. In a realistic large-scale application, classical pre-optimisation of this kind would not be feasible.\\

Emulator and hardware results for the QMC-Givens ansatz on LiH, $\ce{BeH2}$ and $\ce{N2}$ are shown in \Cref{fig:vqe_figures}~\textcolor{blue}{a)--c)}, reporting energy error as a function of Hardware Quantum Credits (HQC) cost, which reflects circuit complexity and realistic hardware resource requirements.
The HQC cost of a circuit is defined as
\begin{equation}
    \text{HQC} = \frac{N_\text{1q} + 10N_\text{2q} + 5N_\text{m}}{5000} \ast  s + 5,
    \label{eq:HQC}
\end{equation}
where $N_{1q}$ denotes the number of single-qubit gates, $N_{2q}$ the number of two-qubit gates, $N_{m}$ the number of reset and measurement operations and $s$ the number of shots taken. All HQC costs reported in this manuscript are calculated for 1000 shots per measurement.\\

For all systems, energy error increases with circuit complexity on both hardware and emulator, with minimum error achieved at one Givens rotation for $\ce{LiH}$ and two for $\ce{BeH2}$ and $\ce{N2}$. Ideal statevector simulations show the opposite trend, with error decreasing monotonically as circuit complexity grows. These results demonstrate that quantum noise has a non-negligible impact on chemical energy calculations, as in this regime, a shallow, few-parameter circuit outperforms a deeper ansatz, even when the latter is theoretically more accurate under noiseless conditions.\\

The H1 device is Quantinuum's first-generation quantum computer. As hardware generations improve and error correction matures, the balance between ansatz complexity and noise susceptibility will shift, with deeper circuits becoming increasingly viable. Nevertheless, designing ans\"atze that appropriately balance expressivity and depth, like the QMC-Givens approach presented here,  will remain important at every stage of hardware development, even in the era of full fault tolerance. 

\subsubsection{Comparison with Unitary Coupled Cluster}

We compare the performance of the QMC-Givens ansatz with a similarly constructed UCC circuit, including single and double excitations selected through a QMC pre-processing step. \Cref{fig:vqe_figures}~\textcolor{blue}{d)--f)} compares the numbers of HQC required by the two ans\"atze for the same number of variational parameters. We find that the QMC-Givens ansatz consistently constructs shallower, lower-cost circuits, which are more amenable to implementation on quantum hardware for similar expressivity.\\

Statevector simulations show comparable energy accuracy for the QMC-Givens and QMC-UCC for $\ce{LiH}$ and $\ce{BeH2}$, as shown in \Cref{fig:vqe_figures}~\textcolor{blue}{g)--i)}. For N$_2$, QMC-UCC achieves better accuracy at high parameter counts, as expected from \Cref{sec:results_qmc}, reflecting the greater expressivity of UCC operators, which allow the preparation of better quality wavefunctions. In practice, the quantum resource requirements of QMC-UCC circuits are significantly higher than QMC-Givens, making this theoretical advantage difficult to realize in realistic conditions. In particular, for the nitrogen molecule the 30-parameter UCCSD ans\"atze shown require an average of $1.3 \times 10^{5}$ HQC for one energy evaluation, while Givens circuits only require $3.5 \times 10^{4}$ HQC. This is comparable to a UCCSD ansatz with only 10 parameters ($3.7 \times 10^{4}$ HQC), which is significantly less accurate, particularly as the N--N bond is stretched (see \Cref{fig:bindingcurve_N2}).\\

\begin{figure}[h]
\centering
\begin{subfigure}[b]{0.48\textwidth}
\centering
\resizebox{0.86\textwidth}{!}{%
\begin{quantikz}[transparent]
\lstick{$q_0:$} & \gate{X} & \qw & \qw & \qw & \gate[6]{G(\theta_4)}\gateinput{0} & \\
\lstick{$q_1:$} & \gate{X} & \gate[2]{G(\theta_0)}\gateinput{0} & \gate[5,label
style={yshift=0.5cm}]{G(\theta_2)}\gateinput{0} & \gate[5,label
style={yshift=-0.5cm}]{G(\theta_3)}\gateinput{0} & \linethrough & \\
\lstick{$q_2:$} & \qw & \gateinput{1} & \gateinput{2} & \gateinput{2} & \gateinput{2} & \\
\lstick{$q_3:$} & \gate{X} & \qw & \linethrough & \gateinput{1} & \gateinput{1} & \\
\lstick{$q_4:$} & \gate{X} & \gate[2]{G(\theta_1)}\gateinput{0} & \gateinput{1} & \linethrough & \linethrough & \\
\lstick{$q_5:$} & \qw & \gateinput{1}& \gateinput{3} & \gateinput{3} & \gateinput{3} &
\end{quantikz}
}%
\caption{QMC-Givens ansatz circuit for LiH with 5 parameters with Givens rotation gates $G(\theta_i)$. Qubit wires passing through a gate box indicate that the gate does not act on the corresponding qubit.}
\label{fig:givens_circ}
\end{subfigure}
\hfill
\begin{subfigure}[b]{0.48\textwidth}
\centering
\resizebox{\textwidth}{!}{%
\begin{quantikz}
\lstick{$q_0:$} & \gate{X} & \qw & \qw & \gate[6]{T(\theta_2)}\gateinput{0} & & & \\
\lstick{$q_1:$} & \gate{X} &  & \gate[5]{T(\theta_1)}\gateinput{0} &  & \gate[2]{T(\theta_3)}\gateinput{0} &\gate[5,label
style={yshift=-0.5cm}]{T(\theta_4)}\gateinput{0}& \\
\lstick{$q_2:$} & \qw &  & \gateinput{2} & \gateinput{2} & \gateinput{1}&\gateinput{2} & \\
\lstick{$q_3:$} & \gate{X} & \qw & \qw &  & & \gateinput{1} & \\
\lstick{$q_4:$} & \gate{X} & \gate[2]{T(\theta_0)}\gateinput{0} & \gateinput{1} & \gateinput{1} &  &\qw & \\
\lstick{$q_5:$} & \qw & \gateinput{1}& \gateinput{3} & \gateinput{3} & &\gateinput{3} &
\end{quantikz}
}%
\caption{QMC-UCC circuit for LiH with 5 parameters. Single excitation gates (with two indices) consist of two Pauli gadgets, and double excitations (with four indices) consist of 8 Pauli gadgets.}
\label{fig:ucc_circ}
\end{subfigure}
\caption{QMC-Givens and QMC-UCC ansatz circuits for LiH with 5 parameters.}
\label{fig:circuits}
\end{figure}
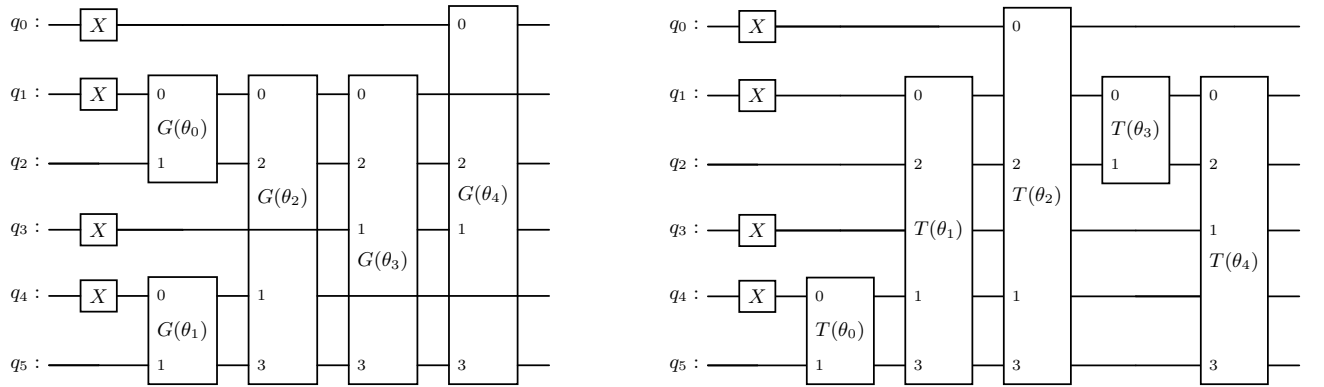

As shown in Figs.~\ref{fig:givens_circ} and \ref{fig:ucc_circ}, this difference in depth stems primarily from the fact that Givens excitations act only on the qubits involved in the electronic excitation, whereas UCC excitations act on all intermediate qubits. Decompositions of Givens rotation gates $G(\theta_i)$ and Pauli gadgets $T(\theta_i)$ are shown in \Cref{fig:givensgates} and \Cref{fig:single_exc_T}, respectively. In this particular example, the UCC operator ordering is optimised for maximum expressivity \cite{Evangelista2019}, while Givens operators are ordered to minimise circuit depth. This is an important consideration in general, but for the simple systems considered here, both orderings can reproduce the same wavefunction. These structural differences render the deeper circuits of QMC-UCC out of reach for current computational capabilities. In noisy settings, QMC-Givens therefore achieves lower energy errors for the same number of parameters.\\

The accuracy gap between the two ansätze may be attributed to the fact that, unlike Givens rotations, UCC operators preserve fermionic anti-commutation relations, producing a better sign structure in the final wavefunction. \Cref{fig:wfn_comp} shows the sign structure of 30-parameter wavefunctions for N$_2$ based on Givens and UCC ans\"atze with single and double excitations. Even where the sign structure agrees within the ansatz space itself, discrepancies emerge in the composite contributions. Correcting these inevitably introduces sign errors elsewhere, making it impossible for the Givens ansatz to fully reproduce a UCC wavefunction.\\
\section{Discussion}

We have introduced a Quantum Monte Carlo approach to tailor Givens rotation based ans\"atze to the capabilities of current quantum devices. Givens circuits are shallower than conventional UCC alternatives, while preserving number symmetry, which is essential for an accurate description of molecular systems. Combined with a classically efficient stochastic pre-screening procedure, the approach yields physically motivated, compact circuits that significantly reduce errors in the presence quantum noise.\\

The optimisation of the variational parameters remains a challenging task, which we accomplished with a non-scalable quantum simulation in this work. However, the QMC prescreening procedure produces coefficients that one can use as a warm start to initialise the parameters in the ansatz. Further research is required to determine how well this procedure scales to larger systems. \\

We implemented the QMC-Givens circuits on Quantinuum System Model H1 to obtain hardware estimates of the ground state energy for different molecular systems. In all cases, the shallowest circuits obtained by QMC-Givens consistently outperform more complex alternatives in the presence of noise. The approach is also significantly cheaper to implement than equivalent UCC circuits, increasing the number of parameters that can be meaningfully encoded within the coherence time of a given device.\\

We show that QMC-Givens wavefunctions are systematically improvable, with more parameters leading to better energies in noiseless conditions. This enables a tunable trade-off between expressivity and circuit depth, tailored to any target device. This flexibility is crucial to the effective use of noisy quantum devices and will remain important even in the fully fault-tolerant regime, as a pathway to optimise resource consumption for a target accuracy.\\

\section{Methods}\label{sec:methods}

\subsection{Givens rotation gates}\label{sec:givens}
An efficient way to generate symmetry-preserving trial states is through Givens rotations \cite{Givens58}. These operators conserve Hamming weight: when applied, they only generate new states with the same number of 1 and 0 qubits. In the Jordan-Wigner encoding, this corresponds to preserving the particle number, so in this sense, they act like excitation operators. The single excitation Givens rotation matrix is given by 

\begin{align} \label{eq:givens1}
G_{1}(\theta) = 
\begin{bmatrix}
    1 & 0 & 0 & 0 \\
    0 & \cos\frac{\theta }{2}& -\sin\frac{\theta }{2}& 0 \\
    0 & \sin\frac{\theta }{2} & \cos\frac{\theta }{2} & 0 \\
    0 & 0 & 0 & 1
\end{bmatrix}
\begin{matrix}
    \ket{00} \\ \ket{01} \\ \ket{10} \\ \ket{11}
\end{matrix}\;.
\end{align}

The matrix only performs a rotation in the subspace of one-particle states, thereby generating a linear combination of the ground state and its excited state. Similar double, triple, etc.~excitation matrices can be found, with efficient quantum circuit implementations available \cite{Magoulas2023,Magoulas2023b}. Examples for the simplest Givens rotations are given in \Cref{fig:givensgates}.\\

Givens gates can also be controlled, with the controlled single excitation Givens rotation matrix given by

\begin{align} \label{eq:givens2}
CG_1(\theta) = 
\begin{bmatrix}
    1 & 0 & 0 & 0 & 0 & 0 & 0 & 0 \\
    0 & 1 & 0 & 0 & 0 & 0 & 0 & 0 \\
    0 & 0 & 1 & 0 & 0 & 0 & 0 & 0 \\
    0 & 0 & 0 & 1 & 0 & 0 & 0 & 0 \\
    0 & 0 & 0 & 0 & 1 & 0 & 0 & 0 \\
    0 & 0 & 0 & 0 & 0 & \cos\frac{\theta}{2}& -\sin\frac{\theta}{2} & 0 \\
    0 & 0 & 0 & 0 & 0 & \sin\frac{\theta}{2} & \cos\frac{\theta}{2} & 0 \\
    0 & 0 & 0 & 0 & 0 & 0 & 0 & 1 \\
\end{bmatrix}
\begin{matrix}
    \ket{000} \\ \ket{001} \\ \ket{010} \\ \ket{011} \\
    \ket{100} \\ \ket{101} \\ \ket{110} \\ \ket{111}
\end{matrix}
\end{align}

which applies the rotation only when the control qubit is equal to $\ket{1}$. \textcite{Arrazola22} showed that such controlled single excitation gates are universal, allowing any particle-conserving unitary to be expressed as a product of them.

\subsection{Controlled Givens Ansatz}\label{sec:controlled_givens}
Given a configuration interaction (CI) expansion,
\begin{equation}
    \ket{\Psi} = \sum_i C_i \ket{\Phi_i}
\end{equation}
any excited determinant in it can be prepared using a controlled Givens rotation. \textcite{Arrazola22} described a general method, which we summarize below, to prepare an arbitrary wavefunction from reference state $\ket{\Phi_0}$ using Givens rotations of the type defined in \ref{sec:givens}.
In the following, we label $c_i = \cos(\theta_i)$ and $s_i = \sin(\theta_i)$.
 \begin{enumerate}
        \item Apply a Givens rotation with angle $\theta_0$ in the subspace spanned by $\{\ket{\Phi_0}, \ket{\Phi_1}\}$:
        \begin{align*}
            \ket{\Phi_0} \xrightarrow{} c_0 \ket{\Phi_0} + s_0 \ket{\Phi_1}.
        \end{align*}
        
        \item Apply a (multi-)controlled excitation with angle $\theta_1$ in the subspace spanned by $\{\ket{\Phi_0}, \ket{\Phi_2}\}$:
        \begin{align*}
          c_0 \ket{\Phi_0} + s_0 \ket{\Phi_1} \xrightarrow{} c_0 c_1 \ket{\Phi_0} + s_0 \ket{\Phi_1} + c_0 s_1 \ket{\Phi_2} . 
        \end{align*}

        \item Repeat this process for the remaining states, to obtain a general function for the Givens ansatz:
        \begin{align} \label{eq:ansatz_function}
        \ket{\Psi(\boldsymbol{\theta})} = \prod_{k=0}^{N-1} \cos(\frac{\theta_k}{2}) \ket{\Phi_0} + \sum_{i=0}^{N-1} \sin(\frac{\theta_{i}}{2}) \prod_{j=0}^{i-1} \cos(\frac{\theta_j}{2}) \ket{\Phi_{i+1}}.
        \end{align}
    \end{enumerate}

From step 2 onwards, (multi-)controlled Givens rotations are used to ensure that the excitation is only applied to the reference state, not affecting the determinants prepared previously. Each excitation amplitude is given by $C_i = \prod_{j \neq (i-1)}^{N-1}\cos(\theta_j) \sin(\theta_{i-1})$. For small values of all $\theta$, $C_i \propto \theta_{i-1}$.

\subsection{Uncontrolled Givens Ansatz} \label{sec:unc_givens}
The controlled Givens ansatz described above is easily generalizable and has a well-defined functional form. However, introducing controlled Givens gates increases the circuit depth significantly when decomposed, making it comparable or greater than conventional alternatives like UCC.\\

To make full use of the low circuit depth of the Givens rotations, we introduce an uncontrolled Givens ansatz, in which the controlled operations are replaced by ordinary Givens gates. Removing the controls means that each new gate acts not only on the reference determinant but also on all determinants generated up to that point, producing additional terms in the wavefunction. 
This is analogous to transforming a configuration interaction wavefunction into a coupled-cluster form, and such parametrisations are commonly referred to as qubit coupled cluster \cite{Xia2020,Yordanov2021,Xie2022}.\\

Much like a coupled cluster wavefunction, the uncontrolled Givens ansatz gives rise to composite contributions, which do not correspond directly to parameters in the wavefunction. These terms will have higher-order dependencies on the parameters, while those stemming directly from an excitation will maintain an approximately linear relationship with the rotation angles. This multiplicative structure ensures that truncated Givens wavefunctions, which do not have an excitation gate for every determinant in the space, maintain desirable properties such as size-consistency and size-extensivity.\\

\subsection{Quantum algorithm}

We used the Jordan--Wigner fermion-to-qubit transformation\cite{Jordan1928} to represent fermionic operators and states on the quantum registers. The Hartree--Fock state was prepared with $X$ gates on qubit indices representing occupied spin-orbitals. We performed parameter optimisation with the L-BFGS-B optimiser implemented within scipy \cite{2020SciPy-NMeth}, in a statevector simulation of the ansatz circuits. No parameter optimisation was performed on hardware or in noisy simulations.\\

After expressing the electronic Hamiltonians as a linear combination of Pauli strings -- tensor products of Pauli operators -- we employed a Pauli partitioning strategy to calculate the energies. The partitioning of Pauli operators into commuting, simultaneously measurable groups was performed via the \texttt{pytket} package~\cite{Sivarajah2020} using the heuristic \texttt{LargestFirst} graph colouring method. The Pauli partitioning for all equilibrium geometry Hamiltonians were saved and reused, ensuring consistency and comparability of the different computations. \\

The quantum circuits were constructed within the \texttt{qiskit}~\cite{Qiskit} and \texttt{pytket}~\cite{Sivarajah2020} frameworks. To make the circuits compatible with the hardware requirements we used the compilation procedure from the Quantinuum Nexus platform \cite{quantinuum_nexus} with optimization level $3$, thus compiling circuits to the Quantinuum H1 native gate set $\{R_{xy}(\theta,\phi), R_{zz}(\theta)\}$, consisting of the single-qubit phased $X$ gate and the two-qubit phased $ZZ$ gates.
We performed quantum hardware calculations on Quantinuum H1 QPU as well as simulations on H1 emulators and on noiseless classical simulators.
To characterize the complexity and cost of circuits we employ Quantinuum's definition of Hardware Quantum Credits (HQCs), detailed in \Cref{eq:HQC}. \\

 The number of qubits, Pauli strings and measurement circuits for all experiments are given in \Cref{circuit_resources}.

\begin{table}[H]
\centering
\caption{Number of qubits, Pauli strings and measurement circuits 
for all experiments. The number of measurement circuits is equivalent to the number of commuting Pauli groups.} 
\label{circuit_resources}
\begin{tabular}{P{18mm} P{25mm} P{25mm} P{25mm}}
\toprule
 & \textbf{LiH} & \textbf{BeH$_2$} & \textbf{N$_2$} \\
\midrule 
    Number of qubits & 6 & 8 & 16  \\ 
    Number of Pauli strings & 118 & 193 & 1177 \\
    Number of measurement circuits & 10 & 16 & 33 \\
\bottomrule
\end{tabular}
\end{table}

\paragraph{Dynamical Decoupling.}
When representing the Hartree--Fock state, no further gates than the initial $X$ ones are present in the circuits and some of the qubits are not acted on by any quantum gate. These idle qubits are prone to decoherence effects because of memory error, especially present in trapped ion quantum devices \cite{Dasu2025,QuantinuumH1Datasheet}. To counteract the decoherence, one can add a sequence of gates that overall act as the identity, but that have the effect of suppressing the decoherence. In this way, no qubit is left idle and the circuit unitary remains unchanged. This is known as dynamical decoupling. We employed a simple dynamical decoupling approach where two $X$ gates are added to the idle qubits. We obtained a lower Hartree-Fock energy error when employing this method compared to allowing idle qubits in the circuits.\\

\paragraph{H1 quantum hardware.}
Quantinuum System Model H1 is a generation of quantum computers with a single linear geometry. We used the H1-1 machine \cite{QuantinuumH1-1_2025}, a 20-qubit trapped-ion quantum computer with full qubit connectivity. It belongs to the NISQ era, with two-qubit gate errors of the order of $9.7\times10^{-4}$ and memory error per depth-1 circuit time of $2.2\times10^{-4}$.
The H1-1E emulator is a classical emulation program that incorporates a noise model based on the performance of H1 hardware. It has noise channels for one and two-qubit gate errors, measurement error, crosstalk, preparation error as well as dephasing and spontaneous emission.

\bibliography{References/references}

\section*{Acknowledgements}
The authors thank Quantinuum for hardware access.
L.C.T.\ acknowledges Quantinuum for studentship funding.
M.A.F.\ acknowledges financial support from Peterhouse, Cambridge through a Research Fellowship and Downing College, Cambridge through the Kim and Julianna Silverman Research Fellowship.

\setcounter{page}{1}
\setcounter{section}{0}
\setcounter{figure}{0}
\renewcommand{\thefigure}{S\arabic{figure}}
\renewcommand{\thepage}{\arabic{page}}

\section*{Supplementary Information for ``Shallow Electronic State Preparation for Quantum Chemistry with Quantum Monte Carlo Pre-Selection"}

\subsubsection{QMC-Givens and QMC-UCC circuits building blocks}

\begin{figure}[h]
    \centering
\begin{quantikz}
& \gate[2]{G(\theta)}\gateinput{0}&\ghost{H} \\
&\gateinput{1}& \ghost{H}
\end{quantikz}=\begin{quantikz}
& \ctrl{1} & \gate{R_y(\frac{\theta}{2})} & \targ{}   & \gate{R_y(-\frac{\theta}{2})} & \targ{}   & \ctrl{1} & \\
& \targ{}  &                              & \ctrl{-1} &                               & \ctrl{-1} & \targ{}  &
\end{quantikz}
    \resizebox{\textwidth}{!}{
\begin{quantikz}
& \gate[4]{G(\theta)}\gateinput{0}&\ghost{H} \\
&\gateinput{1}& \ghost{H} \\
& \gateinput{2}& \ghost{H}\\
& \gateinput{3}& \ghost{H}
\end{quantikz}=
\begin{quantikz}
&   & \ctrl{2}  &     \gate{H}      & \ctrl{1}             & \gate{R_y(-\frac{\theta}{8})}            & \ctrl{3}  &           &           & \gate{R_y(-\frac{\theta}{8})}            &           & \targ{}   & \gate{R_y(\frac{\theta}{8})}             &           &           & \ctrl{3}  & \gate{R_y(\frac{\theta}{8})}             & \ctrl{1}  & \targ{}   & \gate{H}            & \ctrl{2}  &           &\\
&           &           &           & \targ{}                        & \gate{R_y(\frac{\theta}{8})}  &           &           & \targ{}   &            \gate{R_y(\frac{\theta}{8})}  & \targ{}   &           &           \gate{R_y(-\frac{\theta}{8})} &  \targ{}   &   &                  &            \gate{R_y(-\frac{\theta}{8})} & \targ{}   &           &           &           &          & \\
&  \ctrl{1}   &  \targ{}        &                      & \ctrl{1}            &           &           &           &           &           &            \ctrl{-1} & \ctrl{-2} &                      &           &           &           &           &                      & \ctrl{-2}           &           & \targ{}   & \ctrl{1} & \\
&  \targ{}    &       & \gate{H}  &            \targ{}             &           & \targ{}   & \gate{H}  & \ctrl{-2} &                      &           &           &                     & \ctrl{-2} & \gate{H}  & \targ{}              &           &           &                      & \gate{H}  &           & \targ{} &
\end{quantikz}
}
    \caption{Givens circuits adapted from \cite{Arrazola22}, for single and double excitations.}
    \label{fig:givensgates}
\end{figure}
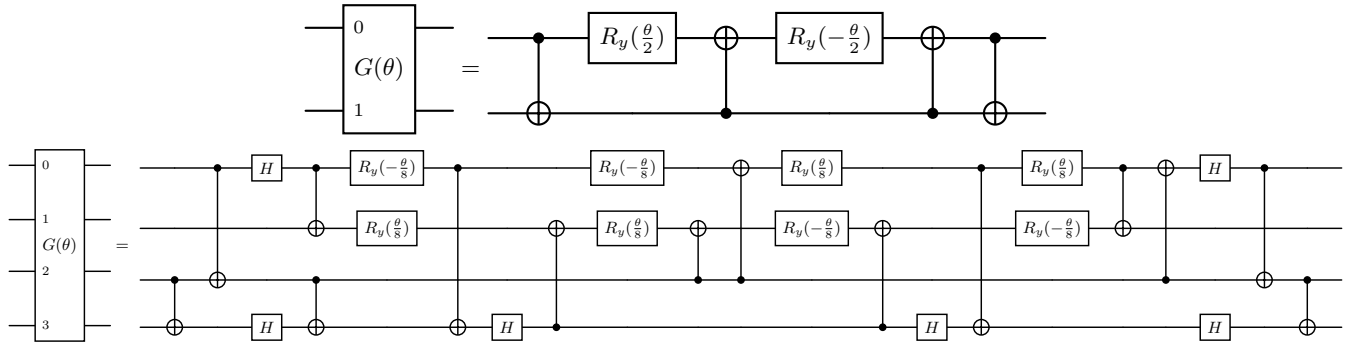

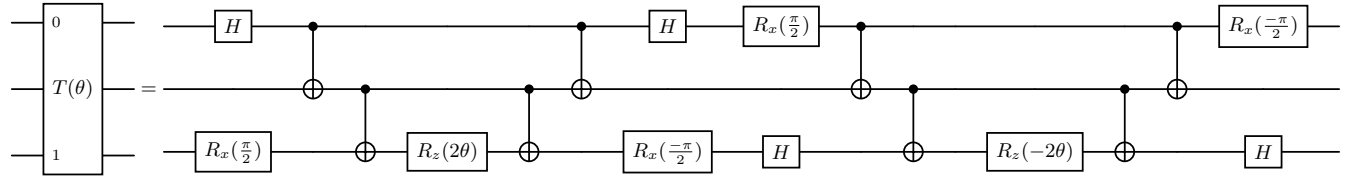
\begin{figure}[h]
    \centering
    \resizebox{\textwidth}{!}{
\begin{quantikz}
& \gate[3]{T(\theta)}\gateinput{0}&\ghost{H} \\
& \ghost{H} & \\
&\gateinput{1}& 
\end{quantikz}=\begin{quantikz}
& \gate{H} & \ctrl{1} &           &                        &           & \ctrl{1} & \gate{H}      & \gate{R_x(\frac{\pi}{2})}  & \ctrl{1} &           &                         &    &       & \ctrl{1} & \gate{R_x(\frac{-\pi}{2})} & \\
&          & \targ{}  & \ctrl{1}  &                        & \ctrl{1}  & \targ{}  &                            &          & \targ{}  & \ctrl{1}  &    &      & \ctrl{1}  & \targ{}  &          &                  \\
&\gate{R_x(\frac{\pi}{2})}  &     & \targ{}     & \gate{R_z(2\theta)}    &    \targ{}       &          & \gate{R_x(\frac{-\pi}{2})}  &             \gate{H} &  & \targ{}        &           & \gate{R_z(-2\theta)}    &    \targ{}       &          & \gate{H} &                 
\end{quantikz}
}
    \caption{Quantum circuit for a single excitation unitary coupled cluster gate. It consists of two Pauli gadgets implementing $e^{2\theta XZY}$ and $e^{-2\theta YZX}$.}
    \label{fig:single_exc_T}
\end{figure}

\subsubsection{Number of shots}

\begin{figure}[h]
    \centering
    \includegraphics[width=0.6\linewidth]{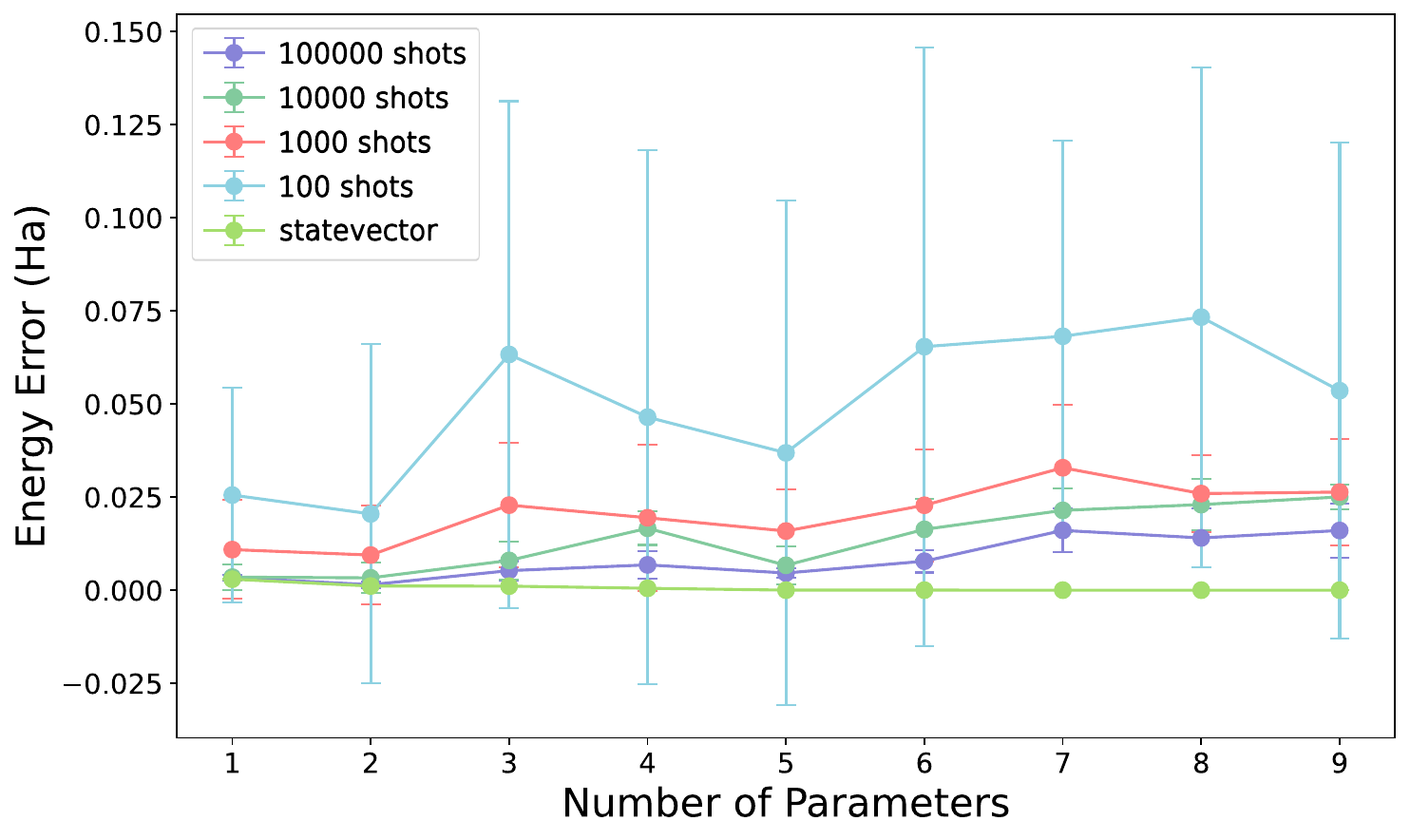}
    \caption{Energy error against number of parameters for \ce{BeH2} QMC-Givens circuits on a noiseless shots simulator with increasing number of shots and on a statevector simulator. Error bars were obtained from 20 repetitions of each experiment.}
    \label{fig:nb_shots_comparison}
\end{figure}

Based on preliminary experiments, varying the number of shots as shown in figure \ref{fig:nb_shots_comparison}, we selected a number of 1000 shots for all experiments. This number offers a balanced trade-off between energy variance and required resources.

\subsubsection{Parameter initialisation}

\begin{figure}[h]
    \centering
    \includegraphics[width=0.6\linewidth]{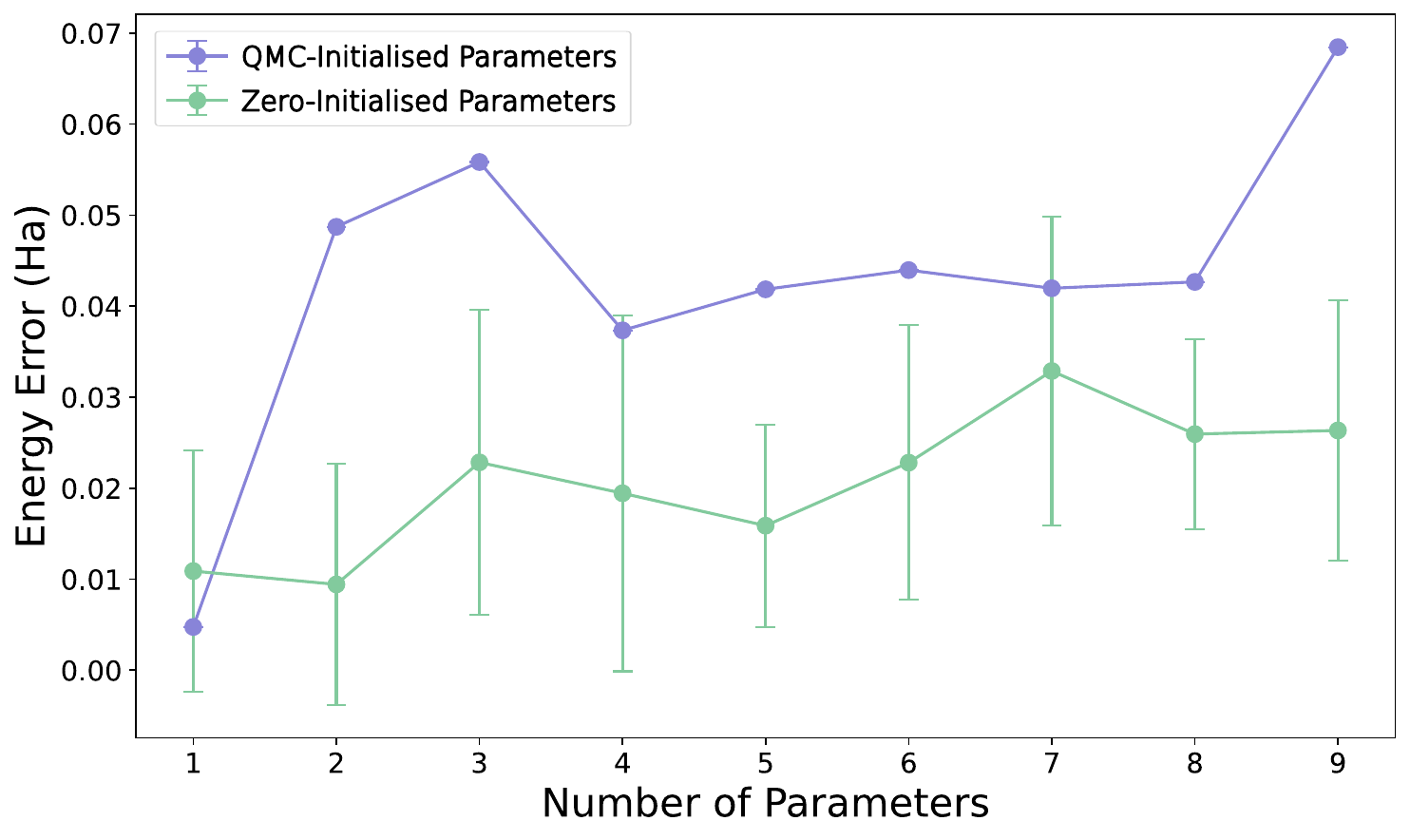}
    \caption{Final-state energy error for BeH2 as a function of the number of variational parameters at 1000 shots, comparing zero-initialized and QMC-initialized parameter strategies.}
    \label{fig:param_mode_comparison}
\end{figure}

 The QMC prescreening procedure produces coefficients for the variational parameters that one can use as a warm start to initialise the parameters before the VQE optimisation. We compared this strategy with a simple zero-initialisation strategy for the parameters, as shown in figure \ref{fig:param_mode_comparison}. These experiments indicate that initialising the parameters at zero is a better strategy, as we obtained a lower energy error for the 1000 shots results when starting from zero-initialised parameters. We maintained the zero-initialisation strategy for all experiments in this work.

\subsubsection{QMC-Givens and QMC-UCC wavefunction comparison}

\begin{figure}[h]
    \centering
    \includegraphics[width=0.8\linewidth]{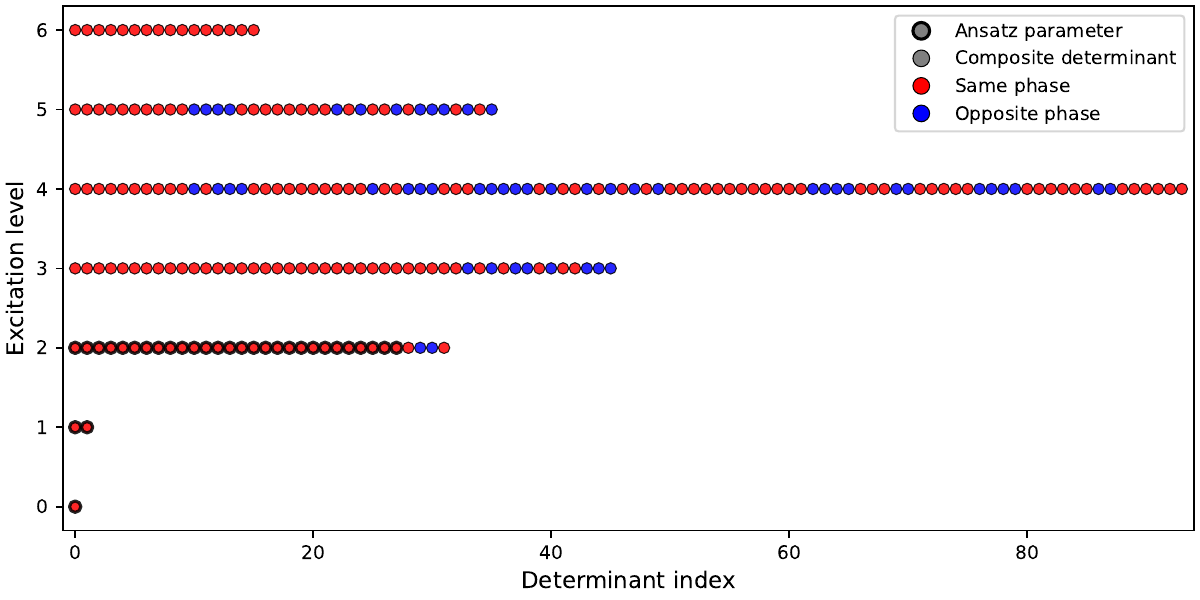}
    \caption{Relative phase of determinants in N$_2$ wavefunctions generated by 30-parameter Givens and UCC ans\"atze with single and double excitations. Both use the same excitations, with corresponding parameters of the same amplitude. The signs of the parameters are chosen to guarantee that
    $\exp(\theta_i \tau_i)\ket{\Phi_0} = G_i(\theta_i')\ket{\Phi_0}$, where $\tau_i$ is the anti-hermitian version of the excitation operator and $G_i$ is the Givens rotation corresponding to the same excitation. The phases contributing to these fixed determinants are marked with bold margins.}
    \label{fig:wfn_comp}
\end{figure}

\end{document}